\documentclass[preprint,floats,eqsecnum,aps,tightenlines,showpacs]{revtex4}
\usepackage{epsfig}

\newcommand{\be}{\begin{equation}}
\newcommand{\ee}{\end{equation}}
\newcommand{\bea}{\begin{eqnarray}}
\newcommand{\eea}{\end{eqnarray}}
\newcommand{\nn}{\nonumber}

\newcommand{\ep}{i\epsilon}
\newcommand{\dq}{\frac{d^4q}{(2\pi)^4}}

\newcommand{\om}{\omega}

\begin{document}

\title{Solving the  Bethe-Salpeter equation for fermion-antifermion pseudoscalar bound state in Minkowski space} 

\author{V.~\v{S}auli}
\affiliation{CFTP and Departamento de F\'{\i}sica,
Instituto Superior T\'ecnico, Av. Rovisco Pais, 1049-001 Lisbon,
Portugal }
\affiliation{Department of Theoretical Physics,
Nuclear Physics Institute, \v{R}e\v{z} near Prague, CZ-25068,
Czech Republic}

\begin{abstract} 
The new method of solution of the Bethe-Salpeter equation for quark-antiquark 
pseudoscalar bound state is proposed.
With the help of integral representation the results are directly obtained in Minkowski space. 
Dressing of Greens functions is naturally considered providing thus the correct inclusion  of the running  coupling constant and  the  quark propagators as well as.   
The first numerical results are presented for a simplified ladder approximation. 
\end{abstract}
\pacs{11.10.St, 11.15.Tk}
\maketitle

\section{Introduction}

Among various approaches used in meson physics the formalism of   Bethe-Salpeter and Dyson-Schwinger equations (DSE's) plays
the traditional and indispensable role.  Bethe-Salpeter equation (BSE) provides
quantum field theoretical starting point to describe hadrons as relativistic bound states of quarks (antiquarks). 
DSE's and BSE's framework has been widely  used in order to obtain the nonperturbative information about the spectra and 
decays of the whole lowest pseudoscalar nonet with emphasizing attention to the  QCD pseudo-Goldstone bosons - the pions
\cite{PSEUDO}.  
With certain success the  formalism
offers a satisfactory window into the 'next scale' meson sector including  also excited and vector,  scalar mesons \cite{SCALARY}. Within formalism, the  mesons electromagnetic form factors have been calculated at spacelike regime of momenta \cite{FORMFAKTORY}.
 
When dealing with bound states  composed from the light quarks then there is unavoidable  
necessity of the full use of BSE covariant framework. The nonperturbative knowledge of the Greens function composing the kernel of BSE is  required.
 Very often, the problem is   solved in the Euclidean space where it is  more tractable 
since there is no singularities in Green functions there.
The  physical amplitudes can be then obtained by continuation to the Minkowski space.
Note, the extraction of the mass spectra is already a complicated task \cite{BHKRW2007},
while there is a practical absence  of an analytical continuation of  Euclidean space form factors.

When dealing with the heavy quarkonia or  with the mesons composed from the mix like $B_c$ 
meson (the state measured in the Fermilab
Tevatron by CDF Collaboration) some simplifying  approximations are possible. 
Different approaches have been developed to reduce computational complexity of full 4-dim BSE. The so called instantaneous approximation \cite{INSTA} 
or quasi-potential approximation \cite{QUASI} can reduce the 4-dim BSE to the 3-dim equation in a Lorentz covariant manner. In practise, such 3-dim equations are more tractable since the solution is less involved, especially when one tricky exploits a considerable freedom in performing three dimensional reduction. Note, contrary to ladder BSE, these equations also reduce to the Schrodinger equation of non-relativistic Heavy Meson Effective Theory and nonrelativistic QCD \cite{HEAVY}. However, the interaction kernels of the reducted equations
often represents the input based on economical model phenomenology and the connection 
to the underlying theory (QCD) is less clear (if not abandoned from the beginning).

In this paper, we  extend the method of solution of the full 4-dim BSE, originally developed for  the pure scalar theories \cite{NAKAN},\cite{KUSIWI},\cite{SAUADA2003},     
 to the theories with non-trivial spin degrees of freedom. Within  a certain assumptions 
on the functional form of Greens functions we develop the method of direct Minkowski space BSE solution  in its original manifestly Lorentz covariant (four-dimensional) manner. 
In order to make this paper as self-contained as possible, we will supply some basic  facts
concerning the BSE approach to relativistic meson bound states bellow here.

The crucial step 
to derive the homogeneous BSE for bound state is the assumptions that the bound state reflects itself in  a pole 
in the four-point Green's function for an on-shell momentum $P$, $P^2=M_j^2$
\be
G^{(4)}(p,p',P)=\sum_j\frac{-i}{(2\pi)^4}\frac{\psi_j(p,P_{os})\bar{\psi_j}(p',P_{os})}{2E_{p_j}(P^0-E_{p_j}+\ep)}+{\rm reg. term}     
\ee 
where $E_{p_j}=\sqrt{\vec{p}^2+M_j^2}$, $M_j$ is a positive mass of the bound state characterized by BS wave function
$\psi_j$ within a set of quantum numbers $j$ . 
Then the BSE can be conventionally written in momentum space like:
\bea
S_1^{-1}(p_+,P)\psi(p,P)S_2^{-1}(p_-,P)&&=-i\int\frac{d^4k}{(2\pi)^4}V(p,k,P)\psi(p,P)\, ,
 \\
p_+&&=p+\alpha P \, ,
\nn \\
p_-&&=p-(1-\alpha)P \, ,
\nn
\eea
or equivalently written in the term of BS vertex function $\Gamma$:
\bea  \label{wakantanka}
\Gamma(p,P)&=&-i\int\frac{d^4k}{(2\pi)^4}V(p,k,P)S_1(k_+,P)\Gamma(p,P)S_2(k_-,P) \, ,
\eea
where we suppress all the Dirac, flavor and Lorentz indices and $\alpha\in(0,1)$. 
The function $V $ represents  two body irreducible interaction kernel and
 $S_i ,i=1,2$ are the dressed propagators of the
constituents. The free propagators reads
\be
S_i^0(p)=\frac{\not p+m_i}{p^2-m^2+\ep},
\ee

Concerning the  solution of BSE (\ref{wakantanka}) for pseudoscalar meson, it  has the generic 
form \cite{LEW}:    
\be \label{gen.form}
\Gamma(q,P)=\gamma_5[\Gamma_A+\Gamma_Bq.P\not\!q+\Gamma_C\not\!P+
\Gamma_D\not\!q\not\!P+  \Gamma_E\not\!P\not\!q]
\ee
where
$ \Gamma_i, i=A,B,C,D,E $  are the scalar functions of their arguments
$ P,q$. If  the bound state has well defined charge parity, say 
${\cal{C}}=1$ then they are even function of  $q.P$ and 
$\Gamma_D=-\Gamma_E$. 

As was already discussed in the paper \cite{MUNCZEK}, the dominant contribution to the BSE calculation for pseudoscalar meson comes namely from the first term in Eq.  (\ref{gen.form}).
Within 15 percentage accuracy it is already true for the light  mesons ($\pi,K,\eta$) while in the case of heavy meson pseudoscalar ground state $(\eta_c, \eta_b)$ the contributions due to the other tensor components in Eq.  (\ref{gen.form}) are even more negligible.
Hence, in  this  stage of our Minkowski calculation  we also approximate our solution by 
taking  $\Gamma=\gamma_5\Gamma_A$.

The interaction kernel is approximated by the dressed gluon propagator
with  the interaction gluon-quark-antiquark vertices  taken in their bare forms,
thus  we can write
\be \label{landau}
V(p,q,P)=g^2(\kappa) D_{\mu\nu}(p-q,\kappa)\gamma^{\nu}\otimes\gamma^{\mu} \, ,
\ee
where the full gluon propagator is renormalized at the  scale $\kappa$  and the
effective strong running coupling are related through the following equations:  
\bea  \label{gluon}
g^2(\kappa)D_{\mu\nu}(l,\kappa)&&=
\alpha_s(l,\kappa)\frac{ P^T_{\mu\nu}(l)}{l^2+\ep}-\xi g^2(\kappa)\frac{l_{\mu}l_{\nu}}{l^4+\ep}\, ,
\\
\alpha_s(q,\kappa)&&=\frac{g^2(\kappa)}{1-\Pi(q^2,\kappa)}\, ,
\nn \\
P^T_{\mu\nu}(l)&&=-g_{\mu\nu}+\frac{l_{\mu}l_{\nu}}{l^2}\, .
\nn 
\eea

From the  class of $\xi$ linear covariant gauges the Landau gauge $\xi=0$ is employed thorough the presented paper.

In the next Section we will derive the solution  for the dressed ladder approximation of BSE, i.e. the the all propagators are considered as a  dressed ones and no crossed diagrams are taken into the calculation. The BSE for quark-antiquark states has been many times considered  in Euclidean space even beyond ladder approximation. Notably, the importance of dressing the proper vertices in the light quark sector has been already stressed in \cite{ACHJO}, so our approximations is known to be certainly limited. 
Going beyond rainbow ($\gamma_{\mu}$ approximation) is  straightforward  but rather involved
(for comparison, see the SDE Minkowski study published in \cite{SAULIJHEP} and in \cite{SAULI2},
wherein the later study includes the minimal gauge covariant vertex 
instead of bare one). In this paper we prefer 
to explain the computational method instead of making a BSE study  with the most sophisticated 
kernel known in the literature.

The layout of the papers is as follows: In the next section we describe the method of the solution. As an exhibition of the method, the numerical results are presented
in the Section 3.  We conclude  in Section 4.

\section{ Integral representation and the solution of BSE}

In this section we describe our method of the BSE solution in Minkowski space.
It basically  assumes that the various Greens functions that appears in the interaction kernel of BSE 
can be written as a weighted integral over the various spectral functions 
(the real distribution) $\rho$.

More explicitly stated, the full quark and gluon (in Landau gauge) propagators are assumed to satisfy
the standard Lehmann representation, which reads
\be \label{srforquark}
S(l)=\int_{0}^{\infty}d \om \frac{\rho_v(\omega)\not l+ \rho_s(\omega)}{l^2-\om+\ep} \, ,
\ee
\be \label{srforgluon}
G_{\mu\nu}(l)=\int_{0}^{\infty}d \om \frac{\rho_g(\omega)}{l^2-\om+\ep}P^T_{\mu\nu}(l) \, ,
\ee

The representations \ref{srforquark}, \ref{srforgluon} are strictly related with the singularity structure  and analytical properties of functions used for. Until now, within a certain limitations,  these integral representations have been used for  nonperturbative evaluation of Greens functions in various models \cite{SAFBP}, noting here, the true analytical structure
of QCD Greens functions is not answered (see also \cite{ALKSME,FISHER1})  with reliability. 
 
Alternatively, the integral representation can be more complicated and generalized to the product of such weighted integrals. In fact, our treatment shall counts more general integral representation for quark propagator like the one  recently considered in the paper \cite{SABIA}. Also the following alternative but notable possibility
for gluon propagator, which read now:
\be \label{duosr}
G_{\mu\nu}(l)=\frac{1}{l^2+\ep}\int_{0}^{\infty}d \om \frac{\hat{\rho}_g(\omega)}{l^2-\om+\ep}P^T_{\mu\nu}(l) \, ,
\ee
will be automatically included in our treatment.
Note, this is effectively the gluon form factor rather then gluon propagator itself which satisfies the appropriate integral representation in this case. Such  expression, motivated by analyticized running coupling method \cite{ANAL},
has been recently considered in the studies of dynamical chiral symmetry breaking in various approximations \cite{PAPA,SABIA, SAKLE}.

Further, here we generalize  the idea of Perturbation Theory Integral Representation (PTIR)
\cite{NAKAN} to our case. The PTIR represents unique integral representaion
for n-point Green function defined by n-leg Feynman integral. The inclusion of dressed propagator representing by the internal lines of (skeleton) Feynman diagram  is selfconsistently  achieved by their own IR (see (\ref{srforgluon},{srforgluon})). 

The generalized  PTIR formula for n-point function in theory involving fields with arbitrary
 spin  is exactly the same as in the original scalar theory considered in \cite{NAKAN}
but  the spectral function receives nontrivial tensor structure. Let us denote 
$\rho(\alpha,x_i)$ such  generalized weight function.
Then, it   can be clearly decomposed into the following sum 
\be
\rho(\alpha,x_i)_{scal.th.}\rightarrow \sum_j \rho_j(\alpha,x_i){\cal{P}}_j 
\ee
where $\alpha,x_i$ represent the set of spectral variables
and  where $j$ runs over the all possible independent 
combinations of Lorentz tensors and Dirac matrices $P_j$.
The function $\rho_j(\alpha,x_i)$ represents just the PTIR weight function of $j-$ form factor
(the scalar function by its definition) since it can clearly written  as a suitable scalar
Feynman integral. Leaving the question of (renormalization) scheme dependence,  we refer 
the reader to Nakanishi book \cite{NAKAN} for a detailed derivation of PTIR. 
 
Let us  apply our idea to the pseudoscalar bound state vertex function.
Recognizing that  the singularity structure (given by denominators)
of the rhs. of the  BSE is the same as in the 
scalar models studied in \cite{KUSIWI,SAUADA2003} the appropriate IR for pseudoscalar bound state vertex function $\Gamma_A(q,P)$ has to read     
\be  \label{repr}
\Gamma_A(q,P)=\int_{0}^{\infty} d\om \int_{-1}^{1}dz 
\frac{\rho_A^{[N]}(\om,z)}
{\left[F(\om,z;P,q)\right]^N}\, ,
\ee
where we have introduced useful abbreviation for the denominator of the IR (\ref{repr})
\be \label{efko}
F(\om,z;P,q)=\om-(q^2+q.Pz+P^2/4)-\ep \, ,
\ee
 where $N$ is an integer free parameter. The same IR could apply for the other components of the full function $\Gamma$ which however will  not be  considered furthermore.

Substituting IR's  (\ref{repr}),(\ref{srforgluon}),(\ref{srforquark}) into the
rhs. of BSE \ref{wakantanka} one can analytically integrate over the loop momenta. Assuming Theorem of 
uniqueness \cite{NAKAN}   we should arrive to the same IR (\ref{repr}) because  of the rhs. of BSE \ref{wakantanka}. The derivation  is presented in the Appendix A for the cases $N=1,2$.

In the other words, we have converted the momentum BSE (with singular kernel)
 to the homogeneous two dimensional integral equation for the real
weight function   $\rho_A^{[N]}(\om,z)$
\be \label{madrid}
\rho^{[N]}_A(\tilde{\om},\tilde{z})=
\int_{0}^{\infty} d\om \int_{-1}^{1}dz
V^{[N]}(\tilde{\om}, \tilde{z};\om,z)\rho^{[N]}_A(\om,z)
\ee
where the kernel  $V^{[N]}(\tilde{\om}, \tilde{z};\om,z)$ is the  regular multivariable function.

The kernel $V^{[N]}$ also automatically  supports the domain $\Omega$ where  the function 
$\rho^{[N]}_A(\om,z)$ is nontrivial. This domain  is always  smaller then the infinity strip $0,\infty\times -1,1$ as is explicitly assumed  by the  integral boundaries of $\om,z$ integrals.
For instance, for  simplest  kernel parametrized by the free gluon propagator and with constituent quarks of mass $m$, then for the flavor singlet meson we have   $\rho^{[N]}_A(\om,z)\neq 0$ only for $\om>m^2$.

In our approach, to solve the momentum  BSE in Minkowski space is equivalent to the  finding of the real solution of the real  integral equation (\ref{madrid}). No special choice of the frame or  has been used. If one needs the resulting vertex function can be  obtained by the numerical integration over the $\rho_N$ in an arbitrary reference frame.


\section{ Numerical Results}

In this section we discuss a numerical solutions of the BSE with a various 
interacting kernels. For this purpose
we will vary the  coupling strength and  the effective gluon mass $m_g$ as well as.
We mainly concern on the range  of binding energy that coincides with the heavy quarkonia - the object of future interest of us.
Also we take a discrete mass $m_g$ such that it runs from zero  to the value of constituent mass of quarks, these
values are expected to be  relevant for the case of the  real gluon propagator (when $m_g$ is  replaced by the continuous spectral variable $\om$
(\ref{srforgluon})).  Thus, in each case, the corresponding gluon density is  $\rho_g(c)=N_g\delta(c-m^2_g)$, which specify the kernel of BSE to be (in Landau gauge):
\be  \label{gluon2}
V(q-p)=g^2
\frac{-g_{\mu\nu}+\frac{(q-p)_{\mu}(q-p)_{\nu}}{(q-p)^2}}
{(q-p)^2-m_g^2+\ep}
\gamma^{\nu}\otimes\gamma^{\mu}
\ee
where the prefactor (including the trace of color matrices)   is  simply
absorbed into the coupling constant. For our actual calculation we have used the bare constituent propagators $S_i(p_i)$ with heavy quark mass $M$ (see the Appendix A for this approximation).

Firstly, we followed the standard procedure: after fixing the bound state mass 
($P^2$) we look for 
the solution by iterating the BSE for spectral function with fixed  
coupling constant $\alpha=g^2/(4\pi)$. 
Very similar to the scalar case \cite{SAUADA2003},
the choice  $N=2$ of the power of $F$ in the IR for the bound state vertex function is preferable one.
It reasonably make provision for a  numerical errors from one side and  the computational obstacles for high $N$ from the other side, noting here that using   $N=1$ is rather unhelpful 
(comparing with massive Wick-Cutkosky model) since we did not find the stable solution
for a wide class of input parameters $g,m_g$ for that choice. 
On the other hand, using the value $N=2$,
 we have found the results for the all possible   interaction kernels considered here.
It also includes the cases with vanishing  $m_g$, which means 
the numerical troubles originally  presented in the  scalar models
\cite{SAUADA2003} are fully overcame here. The details of our numerical treatment are presented in Appendix B. 

As it is  more usually in the non-relativistic case,
we  fix the coupling constant $\alpha=g^2/(4\pi)$ and then look for
the  bound state mass spectrum.
We have found the same result in the both cases, whether we fixed $P$ or $\alpha$ as a first, 
noting  that  in the later case the whole  integration in the kernel $K$ needs to be 
performed at each iteration step, which makes the problem  more computer time consuming.

\begin{figure}
\centerline{  \mbox{\psfig{figure=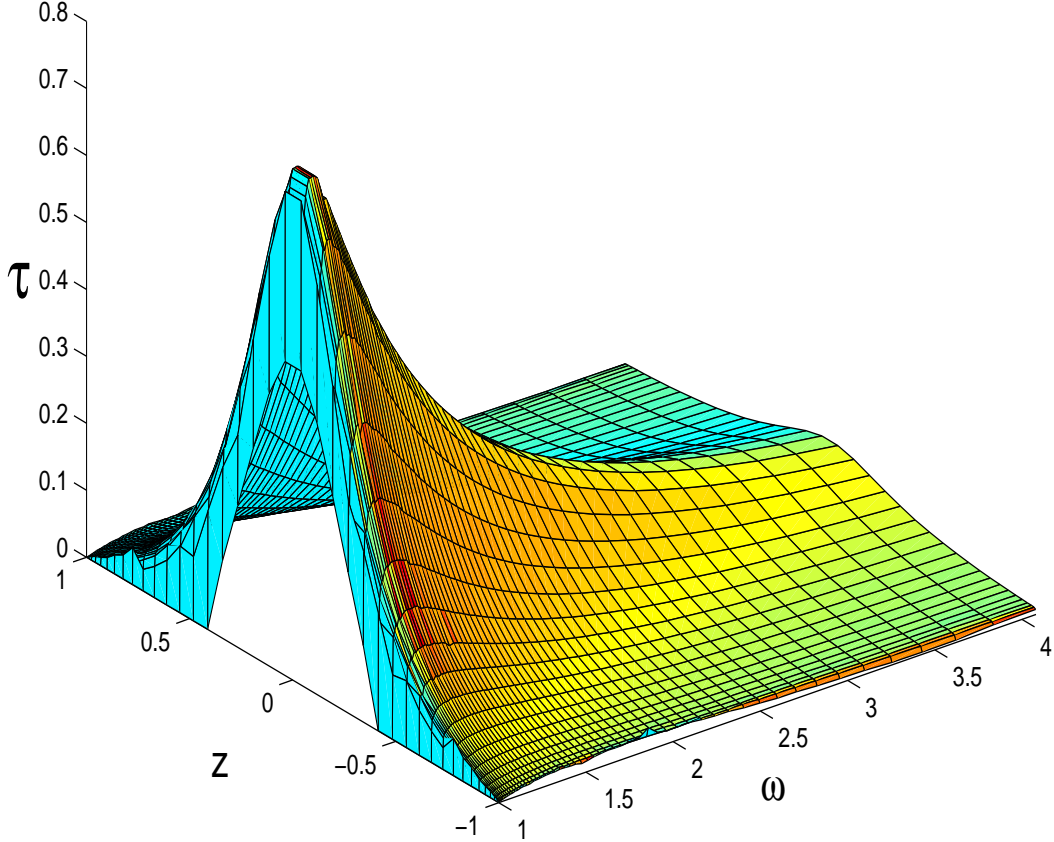,height=14.0truecm,
width=14.0truecm,angle=0}} }
\caption[99]{\label{figBSE} The rescaled weight function 
$\tau=\frac{\rho^{[2]}(\om,z)}{\om^2}$ for the following parameters of the model:
$\eta=0.95, m_g=0.001M_q, \alpha_s=0.666$, the small mass $m_g$ 
 approximates the one gluon exchange interaction kernel. }  
\end{figure}

The obtained  solutions for  
$\alpha's$  for given fractional binding $\eta=\sqrt{P^2}/2m=0.95 $ and various masses $m_g$
are displayed in the Table I.
When we have fixed the gluon mass as  $m_g=0.5 $ and varied the fractional binding
binding $\eta$ we have obtained the spectra as they are displayed in the Table II.

\begin{center}
\small{\begin{tabular}{|c|c|c|c|c|}
\hline \hline
$m_g/m_q $ &  0 $10^{-3}$  & 0.01  & 0.1 & 0.5   \\
\hline 
 $\alpha:$ & 0.666 & 0.669  & 0.745 & 1.029  \\
\hline \hline
\end{tabular}}
\end{center}

\begin{center}
TABLE 1. Coupling constant $\alpha_s=g^2/(4\pi )$  for
several selections of $ m_g/m_q$ ($m_q$ is the 
constituent quark mass) with given fraction of binding $\eta=\sqrt{P^2}/2m=0.95$. 
 \end{center}

\begin{center}
\small{\begin{tabular}{|c|c|c|c|c|}
\hline \hline
$\eta: $ &0.8 &  0.9  & 0.95 & 0.99   \\
\hline 
 $\alpha$ &1.20 & 1.12 & 1.03  & 0.816  \\
\hline \hline
\end{tabular}}
\end{center}

\begin{center}
TABLE 2. Coupling $\alpha_s=g^2/(4\pi )$ as a 
function of fraction of binding $\eta=\sqrt{P^2}/2m$ for  
exchanged massive gluon with $m_g=0.5m_q, m_q$ is the 
constituent quark mass. 
\end{center}

For the illustration the  weight function $\tilde{\rho}^{[2]}$  is
displayed in Fig. (\ref{figBSE}).

\section{ Summary and  Conclusions}
 
The main result of this  paper is  development of   technical framework  
for the solution of Bethe-Salpeter equation in Minkowski space.
To obtain the spectrum, no preferred reference frame is needed and the wave function 
can be obtained in an arbitrary frame (without numerical boosting) by simple integration of the weight function.

The treatment is based on the usage  Integral representation for Green`s function
of given theory, including the bound state vertices themselves. 
The method has been explained and checked numerically  
on the samples of fermion-antifermion pseudoscalar bound state.
It was shown that the momentum space BSE can be converted to the 
real equation for the real weight functions $\rho$ which is easily solvable numerically.
The main motivation  of the author was to develop the practical  tool that respect  
the self-consistence  of Schwinger-Dyson and Bethe-Salpeter equations.
Generalizing this study to the other mesons (vectors, scalars),  considering more general flavor (izospin) structure with the simultaneous improvement  of the approximations (correctly dressed gluon propagator, dressed vertices, etc.) will be a basic steps towards the fully Lorentz covariant description of plethora of  transitions and form factors in timelike region of fourmomenta (Minkowski space).

\appendix
\section{Kernel Functions} 

Writing explicitly Dirac indices the BSE for quark-antiquark bound state reads
\begin{equation} \label{BSE}
\Gamma(q,P)_{\om\rho}=i\int \dq
S(q+P/2)_{\beta\gamma}\Gamma(q,P)_{\gamma\gamma'}
S(q-P/2)_{\gamma'\beta'} V_{\om\beta\beta'\rho}(q,k;P)
\end{equation}
where the Lorentz indices of the vertex function has not been specified.

In our approximation the integral representation for pseudoscalar bound state vertex    
is
\be  \label{reprnul}
\Gamma(q,P)^{\alpha\beta}=\gamma_5^{\alpha\beta}\int_{0}^{\infty} d\om \int_{-1}^{1}dz 
\frac{\rho^{[N]}(\om,z)}
{\left[F(\om,z;P,q)\right]^N}
\ee
and the generalized kernel for  ladder approximation  of the BSE reads
\be \label{irforker}
V_{\alpha\beta\gamma\delta}=g^2\int_{0}^{\infty}d c
\frac{-g_{\mu\nu}+\frac{(q-p)_{\mu}(q-p)_{\nu}}{(q-p)^2}}
{(q-p)^2-c+\ep}\rho_g(c)
\gamma^{\nu}_{\alpha\beta}\gamma^{\mu}_{\gamma\delta}
\ee
where $\alpha,\beta,\gamma,\delta(\mu,\nu)$ 
indices show up the appropriate  Dirac (Lorentz) structure.
In the addition we use the IR for the all functions entering the BSE, this includes
(\ref{reprnul},\ref{irforker}) and the ones for propagators $S(q_\pm)$ (\ref{srforquark}).

For purpose of brevity we will use 
 the following abbreviation for the prefactor:  

\be
\int_{\cal{S}}\equiv
-3\int_{0}^{\infty} d\om \int_{-1}^{+1}dz \int_0^{\infty}dc
 \int_0^{\infty}da\int_0^{\infty} db  \rho^{[N]}(\om,z) g^2\rho_g(c) \, .
\ee

Within this convention the  BSE can be written like

\bea\label{main}
\int d\tilde{\om} d\tilde{z} \frac{\rho^{N}(\tilde{\om},\tilde{z})}
{\left[F(\tilde{\om},\tilde{z};p,P)\right]^N}
=i \int_{\cal{S}} \int\dq 
\frac{\rho_v(a)\rho_v(b)(q^2- P^2/4)-\rho_s(a)\rho_s(b)}
{\left[F(\om,z;q,P)\right]^N D_1 D_2 D_3} \, ,
\eea
where the trace was taken over the Dirac indices (after multiplying by $\gamma_5$)
and
\bea
D_1&=&(q+P/2)^2-a+\ep \, ,
\nn \\
D_2&=&(q-P/2)^2-b+\ep \, ,
\nn \\
D_3&=&(q-p)^2-c+\ep \, .
\eea
In what follows, we will transform the rhs. of BSE (\ref{main}) into 
the integral representation, ie. lhs. of  BSE (\ref{main}).

As a first we use the  following  algebraic identities:
\be
\frac{q^2}{F(\om,z;q,P)}=
\frac{\om-q.Pz-P^2/4}{F(\om,z;q,P)}-1 \, ,
\nn
\ee
\be
\frac{q.P}{D_1D_2}=\frac{1}{2}\left(\frac{1}{D_1}-\frac{1}{D_2}
+\frac{b-a}{D_1D_2}\right) \, ,
\ee
which gives us for the rhs. of  (\ref{main}):
\bea \label{prava2}
&&i \int_{\cal{S}} \int\dq
\left\{\frac{-\frac{z}{2}\rho_v(a)\rho_v(b)}{\left[F(\om,z;q,P)\right]^N D_3}
\left(\frac{1}{D_1}-\frac{1}{D_2}\right)\right.
\nn \\
&+&\frac{(\om-P^2/2+\frac{a-b}{2}z)\rho_v(a)\rho_v(b)-\rho_s(a)\rho_s(b)}
{\left[F(\om,z;q,P)\right]^N D_1 D_2 D_3}
-\frac{\rho_v(a)\rho_v(b)}{\left[F(\om,z;q,P)\right]^{N-1}D_1 D_2 D_3}
\eea

In addition we will use Feynman parametrization, starting with the first term in  the Rel. (\ref{prava2}) we can get:
\bea \label{match}
&&\frac{1}{\left[F(\om,z;q,P)\right]^N }
\left(\frac{1}{D_1}-\frac{1}{D_2}\right)=
(-)^N\int_0^1dx\frac{\Gamma(N+1)x^{N-1}}
{\Gamma(N)}
\nn \\
&&\left\{\left[q^2+q.P(zx-(1-x))+P^2/4-\om x-a(1-x)\right]^{-N-1}\right.
\nn \\
&&-\left.\left[q^2+q.P(zx+(1-x))+P^2/4-\om x-b(1-x)\right]^{-N-1}\right\}\, .
\eea 
Substituting (\ref{match}) back to the relation (\ref{prava2}),
using the Feynman variable $y$ for matching with the scalar propagator $D_3$  and 
integrating over the four-momentum $q$ we get for the first line in 
(\ref{prava2}) the following expression:
\bea   \label{racek}
&&(-)^{N-1}\int_{\cal{S}}\frac{\rho_v(a)\rho_v(b)}{2(4\pi)^2}
\int_0^1 dx\int_0^1dy y^Nx^{1-N}z
\nn \\
&&\left\{\left[\frac{P^2}{4}\left[y-y^2(x(1+z)-1)^2\right]
+q^2(1-y)y+P.py(1-y)(x(1+z)-1)-U(a)\right]^{-N}\right.
\nn \\
&&-\left.\left[\frac{P^2}{4}\left[y-y^2(x(z-1)+1)^2\right]
+q^2(1-y)y+P.py(1-y)(x(z-1)+1)-U(b)\right]^{-N}\right\}
\eea
where we have labeled
\be
U(a)=\left[\om x+a(1-x)\right]y+c(1-y) \, .
\ee
 Taking the substitution $x\rightarrow \tilde{z}$ such that
$\tilde{z}=x(1+z)-1$ in the firts line inside the braces we can  write (including prefactors):
\bea
(-)\frac{\int_{\cal{S}}}{(4\pi)^2}
\int_0^1 dy\int_{-1}^z d\tilde{z}
\left(\frac{1+\tilde{z}}{1+z}\right)^{N-1}
\frac{z}{1+z}\frac{\rho_v(a)\rho_v(b)}{2(1-y)^N
\left[F(\tilde{\Omega},\tilde{z};p,P)\right]^N} \, ,
\eea
where $F$ is defined by (\ref{efko}) and 
\be \label{konvert}
\tilde{\Omega}=\frac{\left(\om\frac{1+\tilde{z}}{1+z}
+a\frac{z-\tilde{z}}{1+z}\right)y+c(1-y)
-\frac{P^2}{4}y^2(1-\tilde{z}^2)}
{y(1-y)} \, .
\ee
Introducing the identity 
\be
1=\int_0^{\infty}d \tilde{\om} \delta(\tilde{\om}-\tilde{\Omega})
\ee
into the Rel. (\ref{konvert}), changing the integral ordering and
integrating over the Feynman variable $y$ we can obtain the desired expression:
\bea   \label{result1}
&&\int_0^{\infty} d \tilde{\om}\int_{-1}^1 d\tilde{z}
\frac{\chi_1(\tilde{\om},\tilde{z})}
{\left[F(\tilde{\omega},\tilde{z};p,P)\right]^N};
\nn \\
&&\chi_1(\tilde{\om},\tilde{z})=
-\int_{\cal{S}}\frac{T_+^{N-1}}{2(4\pi)^2}
\frac{z\rho_v(a)\rho_v(b)}{1+z}\theta(z-\tilde{z})
\sum_{j=\pm}\frac{y_{aj}\theta(D_a)\theta(y_{aj})\theta(1-y_{aj})}
{(1-y_{aj})^{N-1}\sqrt{D_a}} \, ,
\eea
where $y_{a\pm}$ are the roots of the quadratic equation
\be
y^2A+yB_a+c=0
\ee
with the functions $A, B_a, D_a$ defined like
\bea \label{delfin}
&&A=\tilde{\om}-S; \hspace{1cm}
B_a=(\om-a) T_+ +a-c-\tilde{\om};
\hspace{1cm}D_a=B_a^2-4A c
\nn \\
&&S=(1-\tilde{z}^2)\frac{P^2}{4};\hspace{1cm} 
T_{\pm}=\frac{1\pm\tilde{z}}{1\pm z}.
\eea
Repeating the similar procedure in  the second term in
rel. (\ref{racek}) we can write for this: 
%
\bea \label{result2}
&&\chi_2(\tilde{\om},\tilde{z})=
\int_{\cal{S}}\frac{T_-^{N-1}}{2(4\pi)^2}
\frac{z}{1-z}\rho_v(a)\rho_v(b)\theta(\tilde{z}-z)
\sum_{j=\pm}\frac{y_{bj}\theta{(D_b)}\theta(y_{bj})\theta(1-y_{bj})}
{(1-y_{bj})^{N-1}\sqrt{D_b}},
\nn \\
&&y_{b\pm}=\frac{-B_b\pm\sqrt{D_b}}{2A};\hspace{1cm}B_b=(\om-b) T_- +b-c-\tilde{\om} 
\hspace{1cm}; D_b=B_b^2-4A c 
\eea
where we used the label $\chi_2$ instead of $\chi_1$.

In what follows we transform  the  last term of the second line of Rel. (\ref{prava2})
to the desired form of integral representation (\ref{repr}). For this purpose we basily follow 
the derivation already presented in \cite{SAUADA2003}, since the procedure is exactly the same.
However,  due to notational difference we present all calculational  details bellow.

Let us denote the relevant integral as:
\bea \label{kralik}
&&I=i \dq \frac{1}
{D_1D_2D_3\left[F(\omega,z;q,P)\right]^{N-1}}
\eea

Using the  Feynman parameterization technique we first write 
\bea
  D_1 D_2&=&
 \frac{1}{2} \int\limits_{-1}^1 
 \frac{d \eta}{[M^2- f(q,P,\eta)- i \epsilon]^2} \, , \nonumber \\
  M^2\equiv \frac{a+b}{2}+ \frac{a-b}{2}\, \eta \,& ;&f(q,P,\eta)=q^2+\eta \, q.P +P^2/4 \, .
\eea
Next the denominator of IR for bound state vertex is added:
\bea
&&\frac{D_1 D_2}{[F(\om,z;q,P)]^{N-1}}=
\frac{\Gamma(N+1)}{2\Gamma(N-1)}\,  \int\limits_{-1}^1 d \eta
\int\limits_0^1 d t 
\frac{(1-t)t^{N-2}}{[R - f(q,P,\tilde{z})- i\epsilon]^{N+1}} \, , \nonumber \\
&&R= \om t + (1-t)M^2 \, , 
\eea
where $	\tilde{z}= t z+ (1-t)\eta$. Now, we match with the function $D_3$ 
and  integrate over the four momentum $ q$. Thus we get 
\bea
&&I= i \,  \int \frac{d^4 q}{(2\pi)^4} 
 \frac{D_1 D_2 D_3}{[F(\om,z;q,P)]^{N-1}} \\
&&\hspace*{2.0truecm}
 =- i \, \frac{\Gamma(N+2)}{2\Gamma(N-1)}\,  \int\limits_{-1}^1 d \eta
\int\limits_0^1 d t\, (1-t)t^{N-2}\int\limits_0^1 d x\, x^{N}\, I_q \, , \nonumber\\
&&I_q= \int \frac{d^4 q}{(2\pi)^4} 
\left[ -q^2+ q \cdot Q - (1-x)p^2- \frac{x}{4}P^2
+(1-x) c+ x R- i\epsilon \right]^{-(N+2)} \nonumber \\
&& \hspace*{2.0truecm}
= \frac{i}{(4\pi)^2}\frac{\Gamma(N)}{\Gamma(N+2)}\frac{1}{x^{N}(1-x)^{N}}\,
\frac{1}{[ \Omega(t) - f(q,P,\tilde{z})- i\epsilon ]^{N}} \nonumber \\
&& \Omega(t)\equiv \frac{R}{1-x}+ \frac{c}{x}- \frac{x}{(1-x)}\, S \, ,
\eea
where $Q= (1-x)p- x \tilde{z} P/2$. The function $S$ is defined by (\ref{delfin}) and note $\tilde{z}$ lies in the interval $<-1,+1>$, $0\leq S < (\sqrt{a}+ \sqrt{b})^2/4$. Interchanging the 
integrals over $\eta$ and $t$ such that : 
\bea && \int\limits_{-1}^{1}d\eta\int\limits_0^1 dt= 
\int\limits_{-1}^{1}d \tilde{z}\left[\int_0^{T_+}\frac {dt}{1-t}\, 
\Theta(z-\tilde{z}) +\int_0^{T_{-}}\frac {dt}{1-t}\, \Theta(\tilde{z}-z)\right] \, , 
 \nonumber\\ 
&&T_\pm= \frac{1 \pm \tilde{z}}{1 \pm z} \quad \quad \mbox{and} \quad \quad 
\tilde{z}= t z+ (1-t)\eta \, , 
\eea 
we finally obtain 
\be
I= \frac{N-1}{2(4\pi)^2}  \int\limits_{-1}^1 d \tilde{z}
 \int\limits_0^1 \frac{d x}{(1-x)^{N}} \sum_{s=\pm} \Theta(s(z-\tilde{z}))
 \int\limits_0^{T_s} \frac{d t\, t^{N-2}}{[F(\Omega(t),\tilde{z};p,P)]^{N}} \, , 
\ee

In addition we make $t-$dependence of $F(\Omega(t),\tilde{z};p,P)$ explicit
\be
 F(\Omega(t),\tilde{z};p,P)= \frac{J(\om,z)}{1-x}\, t+  F(\Omega(0),\tilde{z};p,P) \, .
\nonumber 
\ee
where 
\bea
&&\Omega(t)=\frac{R(t)-S}{1-x}+\frac{c}{x}+S,
\nn \\
&&R(t)=J(\om,z)t+\frac{b+a}{2}-\frac{b-a}{2}\tilde{z},
\nn \\
&&J(\om,z)=\om-\frac{b+a}{2}-\frac{b-a}{2}z \, .
\eea
Integrate over the variable  $t$ we get 
\be
\int \frac{d t\, t^{N-2}}{ F(\Omega(t),\tilde{z};p,P)^{N}}= 
\frac{t^n}{(N-1)\, F(\Omega(0),\tilde{z};p,P)\, [F(\Omega(t),\tilde{z};p,P)]^{N-1}} \, .
\nonumber
\ee
and hence

\bea \label{result3}
I&=&\frac{1}{2(4\pi)^2}
\int_{-1}^{1} d\tilde{z}\int_0^1 dx 
\sum_{s=\pm}\frac{\theta[s(z-\tilde{z})]T_s^{N-1}}
{(1-x)^NF(\Omega(0),\tilde{z};p,P)\left[F(A(T_s),\tilde{z};p,P)\right]^{N-1}} \, .
\nn \\
\eea
In order to separate F's in the denominator  we use the following identity
\bea \label{op}
&&\frac{1}{F(\Omega(0),\tilde{z};p,P)\left[F(\Omega(t),\tilde{z};p,P)\right]^{N-1}}
\nn \\
&&=\frac{1-x}{J(\om,z)T_s}\left[\frac{1}{ F(\Omega(0),\tilde{z};p,P)}-
\frac{1}{F(A(T_s),\tilde{z};p,P)}\right]
\frac{1}{\left[F(A(T_s),\tilde{z};p,P)\right]^{N-2}} \, .
\eea
Note, for a given a given $N$ one can repeat N-1 times this algebra 
until the  power of the last factor vanishes, 
which is just the reason  of introducing the operation (\ref{op}).
After this operation the momentum dependence of the 
denominators at each term turns to be formally the same as  in the desired IR. 
Although it is possible to derive
the appropriate formula for any $N$, it would lead to 
an unpractical expressions (probably it cannot be written in a closed form)
and we rather choose one concrete value of $N$.   
Motivated by  the success of the scalar model studies,  
we take $N=2$ from now. 
Explicitly written we have for I
\bea  \label{res4}
&&\frac{1}{2(4\pi)^2}
\int_{-1}^{1} d\tilde{z}
\int_0^1 dx \sum_{s=\pm}\frac{\theta[s(z-\tilde{z})]}
{J(\om,z)(1-x)}
\left[\frac{1}{ F(\Omega(0),\tilde{z};p,P)}-
\frac{1}{F(A(T_s),\tilde{z};p,P)}\right]
\nn \\
\eea
Integrating per-partes in variable $x$ we get for (\ref{res4}) 
%
\bea \label{resa5}
&&\frac{-1}{2(4\pi)^2}
\int_{-1}^{1} d\tilde{z}
\int_0^1 dx \sum_{s=\pm}\frac{\theta[s(z-\tilde{z})]\ln(1-x)}
{J(\om,z)}
\left[\frac{\frac{d\Omega(0)}{dx}}{ F^2(\Omega(0),\tilde{z};p,P)}-
\frac{\frac{dA(T_s)}{dx}}{F^2(A(T_s),\tilde{z};p,P)}\right] \, .
\nn \\
\eea
Implementing the following identity into the integrand of (\ref{resa5}) 
\be \label{arg}
1=\int_0^{\infty}\delta(\tilde{\om}-\Omega(t))
\ee
into the integrand of (\ref{resa5})  and
changing the order of the integration and 
performing integration in the variable $x$, we arrive to the desired result for the second
term in the second line of the Rel. (\ref{prava2}):
\bea   \label{res5}
&&\int_0^{\infty} d \tilde{\om}\int_{-1}^1 d\tilde{z}
\frac{\chi_3(\tilde{\om},\tilde{z})}
{\left[F(\tilde{\omega},\tilde{z};p,P)\right]^N} \, ,
\nn \\
&& \chi_3(\tilde{\om},\tilde{z})=
\int_{\cal{S}}\frac{\rho_v(a)\rho_v(b)}{2(4\pi)^2}
\sum_{j=\pm}
\left\{\frac{\ln(1-x_j(0))}{J(\om,z)}
\theta[D(0)]\theta[x_{j}(0)]\theta[1-x_{j}(0)]
{\rm sgn}\left[\frac{dA(x_j(0))}{dx_j(0)}\right]\right.
\nn \\
&&-\left.\sum_{s=\pm}\theta[s(z-\tilde{z})]
\frac{\ln(1-x_j(T_s))}{J(\om,z)}
\theta[D(0)]\theta[x_j(T_s)]\theta[1-x_j(T_s)]
{\rm sgn}\left\{E[x_j(T_s)]\right\}\right\}
\eea
where we have included omitted prefactor in (\ref{prava2}) and
 $x_j$ are the roots of the delta function argument in  (\ref{arg})
\bea
x_{\pm}(T)&=&\frac{-B(T)\pm\sqrt{D(T)}}{2A}
\nn \\
 D(T)=B(T)^2-4A c \, \, &&, \, \,
B(T)=R(T)-\tilde{\om}-c
\nn \\
\frac{d\Omega(t)}{dx}=\frac{E(x)}{1-x}
\, \, &&, \, \,
E(x)=\tilde{\om}-S-\frac{c}{x^2}\, . 
\eea

Introducing a bit compact notation for the sum over an arbitrary  function $U$ with
parameter $T$
\be \label{konvence}
\sum_{\cal{T}}U(T)\equiv U(0)-\theta(z-\tilde{z})U(T_+)
-\theta(\tilde{z}-z)U(T_-)
\ee
we can rewrite the above relation (\ref{res5}) like
\be  \label{chi3}
\chi_3(\tilde{\om},\tilde{z})=
\int_{\cal{S}}\frac{\rho_v(a)\rho_v(b)}{2(4\pi)^2}
\sum_{\cal{T}}\sum_{j=\pm}
\frac{\ln(1-x_j(T))}{J(\om,z)}
\theta[D(T)]\theta[x_{j}(T)]\theta[1-x_{j}(T)]
{\rm sgn} E[x_j(T)].
\ee

The first term in the second line 
of Rel.(\ref{prava2}) should be derived can treated in very similar fashion as the previous case,
note only a different power of $F$ in the denominator.
Doing this explicitly and adding  a correct prefactor the appropriate  expression reads: 
\bea \label{chi4}
&&\chi_4(\tilde{\om},\tilde{z})=
\frac{\int_{\cal{S}}}{2(4\pi)^2}
\left[\rho_v(a)\rho_v(b)\left(\om-\frac{P^2}{2}+\frac{a-b}{2}z\right)
-\rho_s(a)\rho_s(b)\right]
\nn \\
&&\sum_{\cal{T}}\sum_{j=\pm}
\frac{\theta[D(T)]\theta[x_{j}(T)]\theta[1-x_{j}(T)]}{J^2(\om,z)}
\left\{\frac{TJ(\om,z)}{(1-x_j(T))|E[x_j(T)]|}
-\ln(1-x_j(T)){\rm sgn} E[x_j(T)]\right\}.
\nn \\
\eea
Assuming the validity of Theorem o the Uniqueness
then the momentum BSE is converted to the equation for the weight function. It reads
\be  \label{eqvahy}
\rho^{[2]}(\tilde{\om},\tilde{z})=
\int_{0}^{\infty} d\om \int_{-1}^{1}dz
V^{[2]}(\tilde{\om}, \tilde{z};\om,z)\rho^{[2]}(\om,z) \, ,
\ee
where the kernel is simply 
given by the sum of contributions derived above
\be
V^{[2]}(\tilde{\om}, \tilde{z};\om,z)=
-3g^2 \int_0^{\infty}dc\int_0^{\infty}da\int_0^{\infty} db  
\rho_g(c)\sum_{i=1}^4\chi_i(\tilde{\om}, \tilde{z};\om,z).
\ee

\begin{center}{\bf Heavy quark approximation -unequal mass case}\end{center}

When the quark is sufficiently heavy (say $ m_q(2GeV)>>\Lambda_{QCD}$) then the approximation of quark mass function by a constant appears to be adequate.
Neglecting the selfenergy correction is equivalent to  use a free heavy quark propagator,
this corresponds with the free particle spectral functions: 
\bea
&&M_{1}\rho_v(a)=\rho_s(a)=M_{1}\delta(a-M_{1}^2) \, ,
\nn \\
&&M_{2}\rho_v(b)=\rho_s(b)=M_{2}\delta(b-M_{2}^2) \, ,
\eea   
where the variable $a$, $b$
distinguish the type of quark from which  the bound 
state is composed. For completeness we write down the kernel explicitly here for this case.  
\be
V^{[2]}(\tilde{\om}, \tilde{z};\om,z)=
\frac{-3g^2}{2(4\pi)^2} \int_0^{\infty}dc
\rho_g(c)\left(\chi_1+\chi_2+
\chi_3+\chi_4\right);
\ee

\bea
&&\chi_1=
-T_+\frac{z\theta(z-\tilde{z})}{1+z}
\sum_{\pm}\frac{y_{a\pm}\theta(D_a)\theta(y_{a\pm})\theta(1-y_{a\pm})}
{(1-y_{a\pm})\sqrt{D_a}},
\nn \\
&&\chi_2=
T_-\frac{z\theta(\tilde{z}-z)}{1-z}
\sum_{\pm}\frac{y_{b\pm}\theta{(D_b)}\theta(y_{b\pm})\theta(1-y_{b\pm})}
{(1-y_{b\pm})\sqrt{D_b}},
\nn \\  
&&\chi_3=
\sum_{\cal{T}}\sum_{j=\pm}
\frac{\ln(1-x_j(T))}{J(\om,z)}
\theta[D(T)]\theta[x_{j}(T)]\theta[1-x_{j}(T)]
{\rm sgn} E[x_j(T)],
\nn \\ 
&&\chi_4=
\left(\om-\frac{P^2}{2}+\frac{M_1^2-M_2^2}{2}z
-M_1M_2\right)
\nn \\
&&\sum_{\cal{T}}\sum_{j=\pm}
\frac{\theta[D(T)]\theta[x_{j}(T)]\theta[1-x_{j}(T)]}{J^2(\om,z)}
\left\{\frac{TJ(\om,z)}{(1-x_j(T))|E[x_j(T)]|}
-\ln(1-x_j(T)){\rm sgn} E[x_{j}(T)]\right\}
\nn \\
\eea
and where the arguments $a,b$ in the functions $x,J,D$ and  must be replaced by the quark masses
$M_1, M_2$.

\begin{center}{\bf Equal mass case}\end{center}

In the case of quarkonia the kernel becomes more symmetric with  respect to the  variable $z$
and the  formula for the kernel further  simplify. 
The function $R$ depends on $z$`s only through the variable $T$
such that
\be
J(\om,z)\rightarrow J=\om-M^2;
\hspace{2cm}
R(T)\rightarrow R(T)=J\,T+M^2
\ee
where $M$ is a common mass $M=M_1=M_2$. The roots 
become
\be
y_{a\pm}  \rightarrow x_{j=\pm}(T_+); \hspace{2cm}
y_{b\pm}  \rightarrow x_{j=\pm}(T_-)\, ,
\ee
in this case and the kernel can be written in  more compact form:
\be
V^{[2]}(\tilde{\om}, \tilde{z};\om,z)=
\frac{-3g^2}{2(4\pi)^2} \int_0^{\infty}dc
\rho_g(c) K^{[2]}(\tilde{\om}, \tilde{z};\om,z,c)
\ee
\bea \label{equalmass}
&&K^{[2]}(\tilde{\om}, \tilde{z};\om,z,c)=
\nn \\
&&\sum_{s=\pm}\sum_{j=\pm}\theta[s(z-\tilde{z})]
\theta[D(T_s)]\theta[x_{j}(T_s)]\theta[1-x_{j}(T_s)]
\frac{-szT_s x_{j}(T_s)}{(1+sz) \sqrt{D(T_s)}(1-x_{j}(T_s))}
\nn \\
&&+\sum_{\cal{T}}\sum_{j=\pm}
\theta[D(T)]\theta[x_{j}(T)]\theta[1-x_{j}(T)]
\nn \\
&&\left\{\frac{(1-\frac{P^2}{2J})T}{|E[x_j(T)]|(1-x_j(T))}
+\ln(1-x_j(T)){\rm sgn} E[x_j(T)]
\frac{P^2}{2J^2}\right\}
\eea

\begin{center}{\bf One gluon exchange approximation}\end{center}

In order to consider one gluon  approximation  
we should take the massless limit $c\rightarrow 0$. In addition we also restrict ourselves to 
the equal mass case.
One can easily recognize that  
the root ($x_+=0$) is trivial and associated  contribution in (\ref{equalmass}) vanishes.

For the purpose of brevity we label
\be 
x(T)\equiv x_-(T)=\frac{\tilde{\om}-R(T)}{\tilde{\om}-S}.
\ee

Taking into account the relations
$$\sqrt{D}/x=A, E=A,$$
and doing a little algebra we get for the  kernel:
\bea \label{onefoton}
V^{[2]}_{OGE}(\tilde{\om}, \tilde{z};\om,z)&=&\frac{-3g^2}{(4\pi)^2} 
\left\{\frac{P^2}{4J^2}\theta(\tilde{\om}-m^2)\ln\left(\frac{m^2-S}{\tilde{\om}-S}\right)
\right.
\nn \\
&+&\left.\sum_{s=\pm}\theta[s(z-\tilde{z})]\theta(-B_s)\theta(A+B_s)
\left[\frac{T_s\left(\frac{1}{2}-\frac{P^2}{4J}\right)}{A+B_s}
+\ln\left(1+\frac{B_s}{A}\right)\frac{P^2}{4J^2}\right]\right\}.
\eea
%
For  completeness we review the  complete list of the  functions at this place:
\bea
&&A=\tilde{\om}-S,
\hspace{1cm}
J=\om-M^2
\hspace{1cm}, S=(1-\tilde{z}^2)\frac{P^2}{4}
\nn \\
&&B_s=(\om-M^2)T_s+M^2-\tilde{\om}
,\hspace{1cm} 
T_{\pm}=\frac{1\pm\tilde{z}}{1\pm z}.
\eea 

\begin{center}{\bf Case N=1 for}\end{center} 

Repeating the derivation for the parameter $N=1$ we should obtain the following homogeneous equation
\be  
\rho^{[1]}(\tilde{\om},\tilde{z})=
\int_{0}^{\infty} d\om \int_{-1}^{1}dz
V^{[1]}(\tilde{\om}, \tilde{z};\om,z)\rho^{[1]}(\om,z)
\ee
where the kernel is given by the following  expression
\bea
&&V^{[1]}(\tilde{\om}, \tilde{z};\om,z)=\int_S\frac{\sum_{i=1}^4\chi_i}{2(4\pi)^2};
\nn \\
&&\chi_1=
-\rho_v(a)\rho_v(b)\frac{z}{1+z}\theta(z-\tilde{z})
\sum_{j=\pm}\frac{y_{aj}\theta(D_a)\theta(y_{aj})\theta(1-y_{aj})}
{\sqrt{D_a}},
\nn \\
&&\chi_2=
\rho_v(a)\rho_v(b)\frac{z}{1-z}\theta(\tilde{z}-z)
\sum_{j=\pm}\frac{y_{bj}\theta{(D_b)}\theta(y_{bj})\theta(1-y_{bj})}
{\sqrt{D_b}},
\nn \\
&&\chi_3=
\rho_v(a)\rho_v(b)\sum_{j=\pm}
\frac{\theta[D(0)]\theta[x_{j}(0)]\theta[1-x_{j}(0)]}
{E[x_j(0)]},
\nn \\
&&\chi_4=
\sum_{\cal{T}}\sum_{j=\pm}
\frac{\theta[D(T)]\theta[x_{j}(T)]\theta[1-x_{j}(T)]}{|E[x_j(T)]|J(\om,z)}
\left[\rho_v(a)\rho_v(b)\left(\om-\frac{P^2}{2}+\frac{a-b}{2}z\right)
-\rho_s(a)\rho_s(b)\right].
\eea
where we have used the same notations and conventions as in the previous Section.
The appropriate derivation is very straightforward and exactly repeats the steps made for the case with $N=2$, 
hence we make only few comments here.
The relations for $\chi_{1,2}$ are in fact derived in the previous section 
(since for general $N$). The relation for $\chi_4$ is adopted from the paper \cite{SAUADA2003}and remaining function $\chi_3$
follows from the conversion of the term with $F^0=1$, i.e.

\be
\int\frac{d^4q}{(2\pi)^4}\frac{1}{D_1D_2D_3} \, ,
\ee
which is presented in the momentum space BSE in that case.
 The appropriate derivation is in fact  more easy then in the case $N=2$ and the 
appropriate  result represents the basic Perturbation Theory Integral Representation for a scalar triangle with one leg 
momentum constrained such that $P^2<(M_1+M_2)^2$.

\section{ Numerical procedure}

In this section we describe the numerical procedure actually used for obtaining
the spectra of the bound states.
Because of the structure of the integral equations to be solved, the treatment  is similar 
 the procedure used in the papers \cite{KUSIWI,SAUADA2003}, 
however we have used some tricky improvements modification here which enforce the numeric stability when the mass parameter $c$ in gluon IR (or simple vector boson mass $m_c$) is to small compared to the constituent mass. We describe the technical difference in that case.  
As a first  we describe  the numerical  treatment for the case  $c\simeq m$.

The equation (\ref{equalmass}) represents the homogeneous linear integral 
equation which solution  needs to be properly normalized. 
 Adopting the following auxiliary normalization condition  for this purpose:
\be \label{adopt}
1=\int_{-1}^{1} dz\int_{0}^{\infty} d\omega \frac{\rho^{[2]}(\om,z)}{J^2}.
\ee
Then we can find that the equal constituent mass BSE can be transformed
into the inhomogeneous integral equation (\ref{actual})
\be  \label{actual}
\rho^{[2]}(\tilde{\om},\tilde{z})=K^{[2]}_I(\tilde{\om}, \tilde{z})
+\int_{0}^{\infty} d\om \int_{-1}^{1}dz
K^{[2]}_H(\tilde{\om}, \tilde{z};\om,z)\rho^{[2]}(\om,z)
\ee
with the following kernels:
\bea \label{cer1}
K^{[2]}_I(\tilde{\om}, \tilde{z})=
\frac{-3g^2}{(4\pi)^2}\frac{P^2}{4} 
\sum_{j=\pm}\theta[D(0)]\theta[x_{j}(0)]\theta[1-x_{j}(0)]
\ln(1-x_j(0)){\rm sgn} E[x_j(0)],
\eea
\bea \label{cer2}
K^{[2]}_H(\tilde{\om}, \tilde{z};\om,z,c)&=&
\frac{3g^2}{2(4\pi)^2}
\sum_{j=\pm}\theta[D(0)]\theta[x_{j}(0)]\theta[1-x_{j}(0)]
\ln(1-x_j(0))\frac{{\rm sgn} E[x_j(0)]-1}{J}
\nn \\
&+&\frac{3g^2}{2(4\pi)^2}\sum_{s=\pm}\sum_{j=\pm}\theta[s(z-\tilde{z})]
\theta[D(T_s)]\theta[x_{j}(T_s)]\theta[1-x_{j}(T_s)]
\nn \\
&&\left\{\frac{szT_s x_{j}(T_s)}{(1+sz) \sqrt{D(T_s)}(1-x_{j}(T_s))}+
\right.
\nn \\
&&\left.\frac{\left(1-\frac{P^2}{2J}\right)T_s}{|E[x_j(T_s)]|(1-x_j(T_s))}
+\frac{\ln(1-x_j(T_s))}{J}{\rm sgn} E[x_j(T_s)]
\frac{P^2}{2J}\right\}.
\eea
The meaning  of the all  functions were established in the previous appendix, 
for  convenience of the reader we review them also here:
\bea
&&x_{\pm}(T)=\frac{-B(T)\pm\sqrt{D(T)}}{2A}
\, ; \, \, 
D(T)=B(T)^2-4A m_c \, , 
\nn \\
&&A=\tilde{\om}-S \, ; \,  \, B(T)=R(T)-\tilde{\om}-m_c \, ,
\nn \\  
&&E(x)=\tilde{\om}-\frac{m_c}{x^2}-S 
\, ; \, \,  R(T)=JT+m^2 \, ; \, \,   J=\om-m^2 \, ,
\nn \\
&&J=\om-m^2 \, ; \, \, S=(1-\tilde{z}^2)\frac{P^2}{4} \, ; \, \,
T_{\pm}=\frac{1\pm\tilde{z}}{1\pm z} \, .
\eea
The equation (\ref{actual})   has been actually used for the numerical solution when $m_c/m_q \simeq 1$.
 
The kernel  (\ref{actual}) is free of the running singularities because of the presence of theta functions, however 
in the case of massless gluon kernel we would have the  singularity just  on the boundary. 
There  $J\rightarrow 0$ as $\omega $ approaches the quarks threshold.
We have found that this instability is avoided if we generate the inhomogeneous term 
in a following manner:
We formally  add the zero in the following form 
\be
(f(\tilde{\om},\tilde{z})-f(\tilde{\om},\tilde{z}))\frac{\rho^{[2]}(\om,z)}{\om^2}
\ee
to the right hand side of  the original homogeneous equation Bethe-Salpeter equation for weight function $\rho$.
Then solving the  equation  
\be  \label{actual2}
\rho^{[2]}(\tilde{\om},\tilde{z})=f(\tilde{\om},\tilde{z})+
\int_{0}^{\infty} d\om \int_{-1}^{1}dz
\left[V^{[2]}(\tilde{\om}, \tilde{z};\om,z) -\frac{f(\tilde{\om},\tilde{z})}{\om^2}\right]\rho^{[2]}(\om,z)
\ee
within  the following normalization condition 
\be 
1=\int_{-1}^{1} dz\int_{0}^{\infty} d\omega \frac{\rho^{[2]}(\om,z)}{\om^2}.
\ee
is equivalent to the solution of the  original BSE. 
As a suitable function we choose 
\be
f(\tilde{\om},\tilde{z})=1
\ee
We have found that this method is applicable for any positive value of $m_c^2$, but is slowly convergent when
is used for previously discussed case $m_c\simeq m_q$. 
Note here,  that up to the small numerical error the equation (\ref{actual}) offers
the same spectra for such $\alpha's$ and $m_g's$  where the both equations (\ref{actual}),(\ref{actual2}) 
are numerically stable. 

The  equation (\ref{actual2}) has been solved by the method of the iteration.
If the iterations failed-- measure being both the difference of the rhs. 
and lhs. of the integral equation and deviation of the auxiliary 
normalization integral from a predefined value-- we were changing 
the coupling constant (in the treatment with $P^2$ fixed, otherwise procedure is opposite)
until the solution was found.
For numerical solution we  discretize integration variables  $\omega$ 
and $z$ using Gauss-Legendre quadratures (with suitable mapping from 
$<-1,+1> \rightarrow <\om_{min},\infty >$ for $\om$ ). Equations 
(\ref{actual2}),(\ref{actual}) is solved on the grid of $N=N_z*N_{\om}$ 
points which are spread on the rectangle $(-1,+1)*(\om_{min}, \infty )$. 
The value $\om_{min}$ is  given by the support of the spectral 
function. 
In the all cases we take $ N_{\om}= 2N_z $.
Examples of a numerical convergence for some cases of bound 
states are presented in  Table III. As we can see,
 there is a rather weak dependence of the eigenvalue $\alpha$ on the number of mash points $N_{\om}$.
The last value is calculated from the  weighted average (WA) with $N_{\om}$ be the appropriate weight.

\begin{center}
\small{\begin{tabular}{|c|c|c|c|c|c|}
\hline
\hline
 $N_{\om}:$  &  32  & 40  & 64 & 80 & $WA$ \\
\hline
\hline
$\eta=0.95;\, m_g/M=10^{-3} $ & 0.6611 & 0.6690 & 0.6697 & 0.6734 & 0.669\\
\hline
$\eta=0.95;\, m_g/M=0.5 $ & 1.037 & 1.0259 & 1.0210 & 1.0229 & 1.029\\
\hline
$\eta=0.99;\, m_g/M=0.5 $ & 0.818 & 0.8127 & 0.8158 & 0.8155 & 0.816\\
\hline
\hline
\end{tabular}}
\end{center}

\begin{center}
TABLE 3. The coupling $\alpha_s=g^2/(4\pi) $ for 
 the ladder BSE with fixed ratio  $m_c/M$ as a function of the number of  mesh-points. 
\end{center}

\mbox{ }

\end{document}